# High Temperature Cuprate-Like Superconductivity at Surfaces and Interfaces

J. C. Phillips

Dept. of Physics and Astronomy, Rutgers University, Piscataway, N. J., 08854

**Abstract**

The realization of high-transition-temperature superconductivity (HTSC) confined to nanometre-sized interfaces has been a long-standing goal because of potential applications and the opportunity to study quantum phenomena in reduced dimensions. Here we discuss HTSC at free surfaces, interfaces, and nanoscale cluster surfaces, and show that the percolative self-organized model of bulk HTSC also gives an excellent description of HTSC in quasi-two-dimensional contexts.

Because of enhanced thermal fluctuations most phase transition temperatures $T_c$ decrease with decreasing dimensionality, so that surface and interfacial phase transitions occur at lower $T_c$'s than bulk $T_c$'s. In general this occurs for most superconductive transitions as well, but there are some exceptions, for which a percolative mechanism is appealing [1,2]. Layered cuprates exhibit the largest known superconductive $T_c$'s, with metallic cuprate planes separated by intervening oxide layers, usually doped with interstitial oxygen [3]. Because interplanar stiffness is smaller than the cuprate planar stiffness, pressure enhances interplanar interactions, and most cuprate $T_c$'s increase substantially with pressure. In addition to the cuprates, some marginally stable materials have shown larger $T_c$'s at surfaces, most dramatically $Na_xWO_3$ (Na tungsten bronze), where the bulk $T_c$ = 1.3K increases to 88K at the surface [4-6]. Recently interfacial enhancement of $T_c$ was reported for insulating and overdoped $(La,Sr)_2CuO_4$ [7].

The advantage of percolative mechanisms is that carriers can be localized efficiently at strong interaction centers (for superconductors, centers of strong electron-phonon interactions), while these centers can be connected by highly conductive paths ("wires") where scattering is weak; this consideration motivated the original suggestion [1]. In the cuprates dopants in the insulating oxide layers provide strong electron-phonon interactions, while the cuprate planes themselves function as connectors. However, this mechanism by itself is ineffective, because most carrier paths will simply remain confined to the highly conductive connectors, where the density of electronic states $N(E_F)$ at the Fermi energy $E_F$ is much larger than at the dopants, whose effectiveness is further reduced by a tunneling transfer factor T. Thus a satisfactory model also requires that the connecting cuprate planes be criss-crossed by domain nanowalls. (Such nanodomains are inherently plausible in the cuprates because of their marginally stable pseudoperovskite structures. Early evidence for the existence of cuprate nanodomains [8] has since been supplemented by extensive STM studies [9].) The nanodomain walls eliminate purely planar conductive paths, and force carriers to follow only percolative paths that alternate between strong interaction interlayer dopants and cuprate planar connectors.

This mechanism can be criticized as unlikely, but high temperature superconductivity (HTSC) occurs only rarely. This percolative mechanism can occur if the percolative paths are self-organized. Factors promoting self-organization in the cuprates are marginal lattice stability, which facilitates diffusion of dopants to bridging sites that enable current paths to bypass nanodomain walls, and strong internal electric field in ionic oxides. Percolative mechanisms have gradually attracted the interest of theorists who have provided further support for them [10-12]. Several recent experiments have confirmed that large anisotropy or surface-volume ratios can produce diamagnetic effects at high T suggestive of HTSC at a slightly lower T [13,14].

There are several new marginally stable families of ionic crystals that exhibit many similarities to the cuprates: there are many atoms per unit cell, with tetragonal-orthorhombic lattice instabilities, vicinal antiferromagnetic phases, etc., including $Li_xZrNCl$ [15] and $LaFeAsO_{1-x}F_x$ [16]. Are these similarities accidental, or can the percolative cuprate model explain HTSC in these materials with no additional assumptions? In fact, the percolative model easily explains the similarities, by bringing these new materials into the general framework of self-organized percolative networks. This has already been done for $Li_x(Zr,Hf)NCl$ ($T_c \sim$ 15K-25K)[19], so now a similar (but brief) discussion is given here for the $LaFeAsO_{1-x}F_x$ family ($T_c \sim$ 26K-43K) [18], which is much larger and the subject of hundreds of recent studies.

Unlike effective medium (virtual crystal) theories, the self-organized percolative model makes specific predictions concerning $T_c$ in the fully optimized limit, in other words, it predicts the least upper bound $T_c^{max}$ (X), where X is a percolative configuration coordinate. In the cuprate case X = <R>, where R is the average number of Pauling resonating valence bonds. Although Pauling's resonating valence bond concept has often seemed fuzzy, and has been widely abused in the HTSC context, in the cuprate case it works well with R(Cu) = 2, in other words, Cu is in a 2+ valence state (17,18).

Because these materials are mechanically only marginally stable, they are strongly disordered when doped, and are generally far from optimized with respect to HTSC. Marginal lattice stability determines the overall scale for $T_c$, as the phonon energy shift measured by neutron scattering associated with Jahn-Teller doubling of the unit cell of LO phonons correlates linearly with $T_c^{max}$ in the cuprates [15]. The least upper bound for $T_c$, called $T_c^{max}$, has a strongly percolative character, as it peaks at <R> = 2, where <R> is the average valence number of all the atoms [15,16].

The master function $T_c^{max}$ (<R>) is shown in Fig. 1 for the cuprates; the point for $Li_xHfNCl$ was discussed previously [19], and we now discuss the point for the $LaFeAsO_{1-x}F_x$ family ($T_c$ ~ 26K-43K). Here R(Fe) = 2, just as for Cu, because Fe is in a 2+ valence state [20] in a virtual crystal model. (This model also shows the beginnings of self-organization, in that the average height h of As is found to shift with x [21]; had the calculation been carried out with a large supercell centered on a F dopant, h(As) would have varied with distance from F. This effect alone shows that $N(E_F)$ varies slowly and smoothly with doping, and hence effective medium models cannot explain HTSC. However, from this one should not conclude that electron-phonon interactions do not cause HTSC, as these interactions do set the overall energy scale through Jahn-Teller distortions on and near percolative paths, and these distortions are especially large near dopants.)

When one calculates <R> for undoped LaFeAsO, one obtains <R> = 2.5, which places $LaFeAsO_{1-x}F_x$ ($T_c$ = 26K) very close to $Li_xZrNCl$ on the master curve of Fig. 1, so that the theory appears to succeed effortlessly. However, $T_c$ is maximized at 43K for pressures near 4

GPA [22], an increase of 60%, which is much more than is seen in cuprates. Thus the equilibrium cuprate master function $T_c^{max}$ (<R>) remains valid for the $Li_xZrNCl$ and $LaFeAsO_{1-x}F_x$ families, but the pressure dependence in the latter family is larger than in the cuprates, as Fe-As bonding is more covalent than largely ionic Cu-O bonding [22], due to the larger Pauling electronegativity X differences (X(Fe) = 1.8, X(As) = 2.0, X(Cu) = 1.9, X(O= 3.5)) in the cuprates.

The percolative master function $T_c^{max}$ (<R>) is determined from bulk data on layered crystals, so one can ask whether or not this function can explain trends in $T_c^{max}$ at surfaces and interfaces. Determining <R> at surfaces and interfaces is much more difficult than in the bulk, where it is natural to average R over all atoms, as the self-organized structure is marginally stable overall, and soft modes that are critically bound to percolative superconductive paths should have long wave lengths, as $T_c^{max}$ is still small compared to the melting temperature. However, in layered thin crystalline films with epitaxial interfaces or at doped free surfaces, similar percolative behavior is expected. This turns out to be the case for the $La_2CuO_4$ - overdoped $(La,Sr)_2CuO_4$ interface, where $T_c \sim 50K$, after enhancement by exposure to ozone from ~ 30 K [7]. The ozone enhancement is readily explained by the addition of oxygen dopants, absorbed by $La_2CuO_4$ to give $La_2CuO_{4+\delta}$ with $\delta \sim 0.15$. However, the maximum $T_c$ obtainable in this way is ~ 30K for both layers separately, apparently producing a mystery.

Let us look at this mystery with the master curve. <R> = 16/7 in LCO = 2.28, = (16 - x)/7 in $La_{(2-x)}Sr_xCuO4$, so with x (or $\delta$) = 0.15, <R> = 2.26, and with x = 0.45, <R> = 2.22. This would give a decrease in <R> between x = 0.15 and x = 0.45 of 0.04, and so we get something like $T_c$ = 35K + (0.04/0.28)[150-35]K = 50K. Of course, this is just a plausible guess at the interfacial structure, but the master function has given the trend correctly, not only qualitatively, but even semi-quantitatively (something no other theory has been able to do: virtual crystal theories predict $T_c$ < 1K, from which they erroneously conclude that electron –phonon interactions do not cause HTSC!).

Now we turn to a much more difficult problem, for which the data base is small, but still robust: a surface monolayer of $A_xWO_3$, where A is an alkali metal (Na [4,5] or Cs[6]). While bulk $Na_xWO_3$ exhibits superconductivity only near 1K, here for Na superconductivity appears around 100K; for Cs there are two phase transitions, a bulk one with lower $T_c$ at higher doping, and a re-entrant percolative one with higher $T_c$ at lower doping. Moreover, Na- and Li- (but not K-) doped surfaces of nanoclusters of $WO_3$ embedded in a variety of nanoporous hosts (carbon inverse opal, carbon nanotube paper, or platinum sponge) show diamagnetic anomalies with an onset of 130K [23]. Note that $WO_3$ (with its simpler unit cell, subject only to Jahn-Teller distortions) itself is nonmagnetic. These data can be combined with the master function $T_c^{max}$ (<R>) to construct a model of percolative self-organization at surfaces. To explain $T_c \sim 130K$ one must assume <R> = 2. To model the free surface, one can assume that each surface $W_s$ is connected to an exterior O and back-bonded to an O image, so that $R(W_s) = 2$. The percolative $W_sO$ surface chain then has <R> =2. These surface chains are entropically broken into stress-relieving fragments. Intercalated Li or Na ions connect the chains, thereby increasing their conductivity and their screening of internal ionic fields, just as in the bulk percolative model [1,2]. The embedded clusters are not connected, so the result is "localized non-percolative superconductivity" [23]. In the free surface case [4-6] thermal fluctuations disrupt superconductivity above 100 K. Both of these points are shown in Fig. 1. Considering the rapid progress in nanoscience, it may be possible to obtain similar pairs of points for other HTSC.

In conclusion, the present discussion shows that the original percolative model [1,2] continues to provide an excellent guide to the phenomenology of HTSC, which transcends the otherwise theoretically insuperable barriers of exponentially complex (NPC, non-polynomial complete), aperiodic self-organization commonly encountered not only in HTSC [17], but also in protein science [24,25].

**Figure Caption**

Fig. 1. The master function for HTSC, $T_c^{max}$ (<R>), provides a least upper bound for bulk layered superconductors. This function is based on the percolative model for self-organized HTSC dopant networks [1,2]. The original figure [2] has been modified in [19] to include $Li_xHfNCl$, and here $LaFeAsO_{1-x}F_x$ is added (it is coincident with $Li_xHfNCl$). The triangle above the bulk point represents the highest $T_c$ obtainable under pressure in $LaFeAsO_{1-x}F_x$; it exceeds the predicted upper limit (see text). There are two data points for $A_xWO_3$ (A= Na or Li), the star corresponding to the free surface [4-6], and the rectangle to nanoclusters [23]. The stress-free nature of these ionic dopant networks is compared to the stress-free intermediate phase seen in covalent network glasses [2] for 2.25 <R> < 2.52.

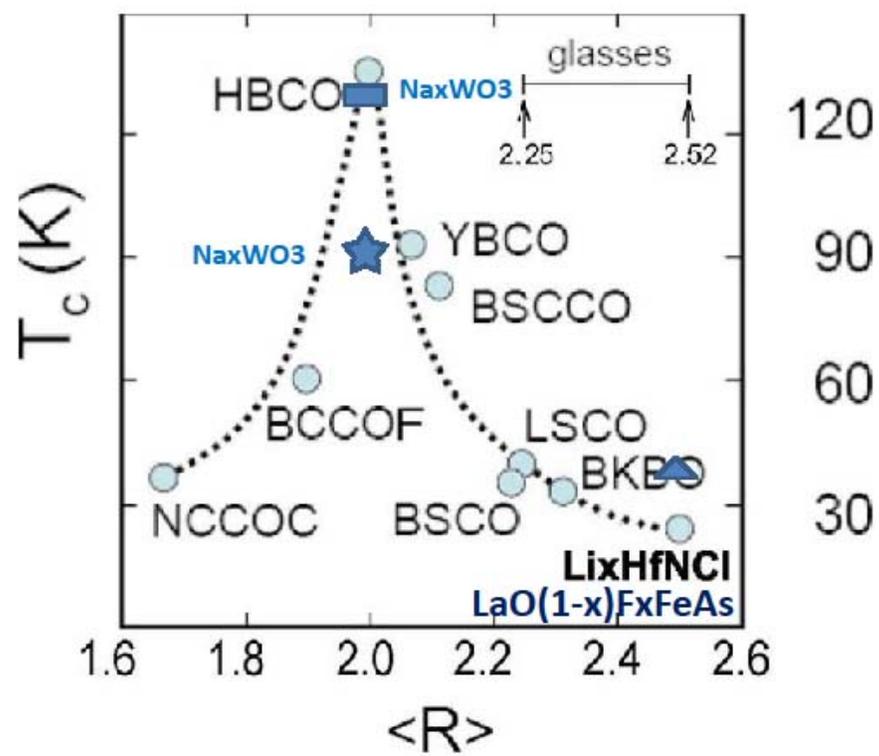